# Valence-band structure of ferromagnetic semiconductor (InGaMn)As

Shinobu Ohya[*], Iriya Muneta, Yufei Xin, Kenta Takata, and Masaaki Tanaka[†]

*Department of Electrical Engineering and Information Systems, The University of Tokyo, 7-3-1 Hongo, Bunkyo-ku, Tokyo, Japan 113-8656*

**Abstract**

To clarify the whole picture of the valence-band structures of prototype ferromagnetic semiconductors (III,Mn)As (III: In and Ga), we perform systematic experiments of the resonant tunneling spectroscopy on $[(In_{0.53}Ga_{0.47})_{1-x}Mn_x]As$ ($x$=0.06-0.15) and $In_{0.87}Mn_{0.13}As$ grown on AlAs/ $In_{0.53}Ga_{0.47}As$:Be/ $p^+$InP(001).  We show that the valence band of $[(In_{0.53}Ga_{0.47})_{1-x}Mn_x]As$ almost remains unchanged from that of the host semiconductor $(In_{0.53}Ga_{0.47})As$, that the Fermi level exists in the band gap, and that the *p-d* exchange splitting in the valence band is negligibly small in (InGaMn)As.   In the $In_{0.87}Mn_{0.13}As$ sample, although the resonant peaks are very weak due to the large strain induced by the lattice mismatch between InP and InMnAs, our results also indicate that the Fermi level exists in the band gap and that the *p-d* exchange splitting in the valence band is negligibly small.   These results are quite similar to those of GaMnAs obtained by the same method, meaning that there are no holes in the valence band, and that the impurity-band holes dominate the transport and magnetism both in the $[(In_{0.53}Ga_{0.47})_{1-x}Mn_x]As$ and $In_{0.87}Mn_{0.13}As$ films.   This band picture of (III,Mn)As is remarkably different from that of II-VI-based diluted magnetic semiconductors.

[*] ohya@cryst.t.u-tokyo.ac.jp
[†] masaaki@ee.t.u-tokyo.ac.jp

## I. INTRODUCTION

Understanding the physics of ferromagnetic semiconductors (FMS) InMnAs[1,2] and GaMnAs[3,4] is one of the central issues in solid state physics, and it is vitally important for developing spin-related functionalities in semiconductors. A lot of unique features found in these material systems are promising for novel semiconductor devices utilizing spin-degrees of freedom.[5-14] However, basic understanding of FMSs, especially on the band structure and the origin of ferromagnetism, is still under debate. Generally, it has been believed that the ferromagnetism in InMnAs and GaMnAs is induced by the spin-polarized valence-band (VB) holes interacting with the localized Mn 3$d$ spin-polarized electrons. It has been assumed that the Fermi level exists in VB which is largely spin-split by the $p$-$d$ exchange interaction and that the VB is merged with the Mn-induced impurity band (IB) as shown in Fig 1 (a).[15-18] This VB conduction picture could explain wide variety of phenomena such as the Curie temperature ($T_C$), magnetic anisotropy, and so on.[19-23] Recently, however, contradictory results were obtained in many experiments on GaMnAs such as the hot-electron photoluminescence, infrared absorption, and magnetic circular dichroism.[24-33] These experiments strongly suggest that the Fermi level exists in IB in the band gap of GaMnAs, as shown in Fig. 1 (b). In this case, the ferromagnetism is stabilized by hopping of the IB holes, which is quite different from the VB conduction picture mentioned above.

There are several problems specific to GaMnAs which prevents a clear determination of the energy-band structure and the Fermi-level position; small energy distance between VB and IB (less than ~100 meV), broadening of the optical spectra due to the small dispersion of IB, and the random distribution of the Mn atoms responsible for the inhomogeneous depletion depth at the surface. These features made the understanding of this material difficult even by the angle-resolved photoemission spectroscopy [34] and scanning tunneling microscopy,[35] which are sometimes very powerful for other material systems.

The resonant tunneling spectroscopy is a very useful method to clarify the VB structure of (III,Mn)As (III=In and Ga). As explained later, the advantage of this method is that we can detect energy bands only with the same wave-function symmetry as that of the $p$-wave functions of the VB holes. Also, it is sensitive to the effective mass of the energy bands. Thus, we can distinguish VB from IB. Furthermore, by carefully analyzing the quantum-well (QW) thickness dependence of the resonant levels, we can avoid the unwanted effect induced at the surface mentioned above. Previously, using this technique, we successfully obtained the VB structure of GaMnAs.[36,37] We found that the Fermi level exists in the band gap, that the $p$-$d$ exchange split in the VB is negligibly small [3-5 meV for the light hole (LH) band], and that the almost unaffected VB of GaAs is still remaining in GaMnAs. These findings indicate that the VB holes do *not* play any crucial role in giving rise to the ferromagnetism in GaMnAs.

From the viewpoint of both physics and applications, it is quite important to investigate the difference of the band structures of FMSs with different host materials. Such studies will give us a guideline on how to design heterostructures to accomplish intended functionalities. Also, it is useful to understand the whole band picture of FMSs. Especially, it is quite important to understand the (III,Mn)As (III=In and Ga) systems, which are the most prototypical FMSs. By increasing the In content in InGaAs, the binding energy of Mn is decreased from ~110 (GaAs) to ~30 meV (InAs).[38] This difference of the binding energies may affect the way of the IB formation or the VB deformation by Mn doping. There have been conflicting reports about the understanding of the VB structure of InMnAs. For example, the cyclotron resonance experiments suggest the VB conduction in InMnAs.[39,40] On the other hand, infrared optical absorption measurements on InMnAs show the double-exchange-like component, suggesting the IB conduction.[41] Therefore, further systematic experiments are needed to investigate the VB structures of (InGaMn)As[42-44] and InMnAs.

**II. EXPERIMENTAL**

Figure 2(a) shows the schematic device structures investigated in this study; mesa diodes with 200 μm in diameter composed of $[(In_{0.53}Ga_{0.47})_{1-x}Mn_x]As$ or $In_{0.87}Mn_{0.13}As$ ($d$ nm) / AlAs (6 nm)/ $In_{0.53}Ga_{0.47}As$:Be (100 nm, Be concentration: $1\times10^{18}$ cm$^{-3}$) grown on $p^+$InP(001). The Mn composition $x$ in $[(In_{0.53}Ga_{0.47})_{1-x}Mn_x]As$ was varied from 0.06 to 0.15. Table I shows the details of the samples used in this study. The $T_C$ values of the (InGaMn)As films (sample A-F) were estimated from the Arrott plot using the data of the magnetic-field dependence of magnetic circular dichroism measured for these samples before etching. The $T_C$ value of the InMnAs sample (sample G), which has a strong in-plane magnetic anisotropy due to the strong compressive strain induced by the lattice mismatch (~3%) between InMnAs and InP, was determined by the planar Hall effect measurements. Some samples were annealed after the growth with the conditions shown in Table I. The (InGaMn)As (or InMnAs) thickness $d$ was precisely controlled by changing the etching time from mesa to mesa.[37] Figure 2(b) shows the schematic VB profiles of these diode devices shown in (a). The thick orange solid curves, thin green solid curves, red dash-dotted lines, and blue lines indicate the heavy-hole (HH) band, light-hole (LH) band, Fermi level, and resonant levels, respectively. Here, the exchange splitting is neglected for simplicity. Due to the large lattice mismatch (~3.6%) between AlAs and InP, the HH and LH bands are largely split in the AlAs layer. We note that this strain also generates the large band offset between (InGaMn)As (or InMnAs) and AlAs, which results in the effective confinement of the VB holes in the (InGaMn)As (or InMnAs) QW. Since (InGaMn)As has a little larger lattice constant than that of InGaAs due to the Mn doping, the HH and LH bands are slightly split in the (InGaMn)As layer. This splitting between HH and LH bands is larger in InMnAs due to the larger strain.  The resonant levels are formed by the VB holes confined by the AlAs barrier and the surface Schottky barrier with the estimated thickness of ~1 nm (depending on the hole concentration) induced by Fermi-level pinning at the surface.  These resonant levels can be detected only when holes are injected from the bottom *p*-type InGaAs:Be layer because the injected holes have small in-plane wave vectors $k_\parallel$ due to the low hole concentration ($1\times10^{18}$ cm$^{-3}$) of the InGaAs:Be electrode.[11]  As shown in Fig. 2(c), the resonant tunneling energies expressed by the blue curves converge to the VB top energy $E_V$ of the bulk (InGaMn)As (or InMnAs) with increasing *d*.  Since the Fermi level corresponds to the zero bias condition, the Fermi level position can be determined by measuring the *d* dependence of the resonant levels.  All the measurements were done in the cryostat cooled at 3.4 - 3.5 K.  Direct current was used to measure the *I-V* characteristics, and $d^2I/dV^2$ values were numerically obtained from the *I-V* data.  The bias polarity is defined by the voltage of the top (InGaMn)As (or InMnAs) electrode with respect to the substrate.

**III. RESULTS**

In Fig. 3, we show the experimental data of the (InGaMn)As and InMnAs resonant tunneling diodes investigated here.  The black solid curves in Fig. 3(a)-(f) correspond to the $d^2I/dV^2$-*V* curves obtained for various *d* at 3.5 K in sample A-F (InGaMnAs samples), respectively.  [In (a), (e), and (f), only a part of the *d* region is shown due to the low yield of our current device fabrication process.]  Oscillations are clearly seen in almost all the curves in Fig. 3(a)-(f).  The oscillation-peak bias voltages get smaller, and the period of the oscillation becomes short with increasing *d* except for the large *d* region in (a) where a very sharp increase in resistance-area (*RA*) occurs in the large *d* region with *d*>16 nm due to the interstitial Mn diffusion to the surface during the low-temperature annealing.  [For details, see Fig. 5(a).]  As mentioned in the Discussion section, these phenomena observed in Fig. 3(a)-(f) are the quite typical QW-thickness dependence of the resonant tunneling effect.  This is the first observation of the resonant tunneling in (InGaMn)As.  Here, HH*n* and LH*n* (*n*: 1,2,3…) represent resonant tunneling through the *n*th level of the HH and LH bands in the (InGaMn)As QW, respectively.  They are assigned by the theoretical analyses mentioned below.  In Fig. 3(b), (c), and (d), we see that the resonant levels are converged at a certain bias voltage corresponding to $E_V$ with increasing *d*.  Also, in Fig. 3(a), (e), and (f), we can see similar tendencies except for the large *d* region with *d*> 16 nm with the sharp *RA* increase in (a).  Considering that Fermi level corresponds to the zero bias condition, we conclude that Fermi level exists in the band gap in all the samples.  We see a specific feature that the converged bias position becomes farther away from the origin with increasing the Mn composition *x* [from Fig. 3(a) to (d)], which means that the Fermi level becomes farther away from $E_V$ with increasing $x$.

If there were large *p-d* exchange splitting, the resonant levels would be split and the splitting energy would increase with increasing $T_C$ or $x$. However, such behavior is not observed in Fig. 3(a)-(f). Therefore, the *p-d* exchange splitting in VB is negligibly small. This result is almost the same as those obtained in GaMnAs.[37] In a resonant tunneling diode with a paramagnetic ZnMnSe QW, clear spin-splitting of the resonant level was observed, and this splitting energy increased with a magnetic field.[45] Our results mean that (III,Mn)As has a quite different band picture from that of II-VI diluted magnetic semiconductors.[46]

The black solid curves in Fig. 3(g) correspond to the $d^2I/dV^2$-$V$ curves in the small bias region ($V$ from -0.1 to 0 V) for various $d$ at 3.5 K in sample D. Three peaks, which do not strongly depend on $d$, are observed at roughly around -0.015, -0.045, and -0.06 V. We note that the similar peaks were observed in the GaMnAs diode devices at around -0.02, -0.05, and -0.065 V. (See the $d$ region from 6 to 16 nm of Fig. 6**c** in Ref. 37.) Since these peaks were enhanced with increasing $x$ or $T_C$, it is expected that these peaks are related to the Mn induced states or a part of the Mn induced IB. These peak positions observed in (InGaMn)As are almost the same as those in (GaMn)As, which means that there is a quite similar electronic structure in the band gap both in GaMnAs and (InGaMn)As. The intensity of these peaks in Fig. 3(g) is much smaller than those of the resonant levels of VB shown in Fig. 3(d), which probably means that these states are mainly composed of the Mn *d* electrons and that they may have a weak component of the *p* wave-function.

The black solid curves in Fig. 3(h) correspond to the $d^2I/dV^2$-$V$ curves obtained for various $d$ at 3.4 K in sample G (InMnAs). When $d$ is less than 6 nm, we see two resonant peaks [HH1 and (HH2 or LH1)] which get close to a certain negative bias voltage with increasing $d$. Therefore, the Fermi level exists in the band gap in $In_{0.87}Mn_{0.13}As$. We do not see any spin-splitting peaks, which would move parallel to the LH peak with increasing $d$ if any.[36] This means that the *p-d* exchange splitting is negligibly small also in $In_{0.87}Mn_{0.13}As$. We see that these peaks disappear when $d$ is larger than 6 nm, which is probably due to the lattice relaxation of the InMnAs layer due to the strong strain. We investigated other InMnAs resonant tunneling diodes with lower Mn concentrations (not shown), but resonant peaks were not clearly observed in them. We think that it is probably due to the stronger strain effect in these InMnAs layers. For more systematic study on VB of InMnAs, it is necessary to use higher-quality strain-free heterostructures.

**IV. COMPARISON WITH THEORETICAL CALCULATIONS**

With a multiband transfer matrix technique developed in ref. 47 with the Luttinger-Kohn 6×6 *k·p* Hamiltonian and the strain Hamiltonian,[16] we calculated the quantum levels in these heterostructures. The Luttinger parameters $\gamma_1$, $\gamma_2$, and $\gamma_3$ of (InGaMn)As (or InMnAs) assumed in our study are 13.4 (20.0), 5.3 (8.5), and 6.0 (9.2), respectively.[48] The shear deformation potential $b$ of (InGaMn)As (or InMnAs) was assumed to be the same as that of InGaAs (InAs); $b$=-1.7 (-1.8).[48] To reproduce the experimental data of the resonant peak bias voltages $V_R$, we assumed the VB diagrams of (InGaMn)As and InMnAs resonant tunneling diodes shown in Fig. 4(a) and (b), respectively. Here, the green (thin black) and orange (thick gray) lines express the LH and HH bands, respectively. Since the $p$-$d$ exchange split was negligibly small, we neglected the $p$-$d$ exchange interaction in our calculations. We divided the surface Schottky barrier region into three equally spaced regions and assumed that the band is flat in each region. $E_F$ is defined as the energy difference between the Fermi level and the HH band at the Γ point in (InGaMn)As (or InMnAs). We used $E_F$ and the deformation anisotropic term[49] $Q_\varepsilon$ as fitting parameters. These parameters were determined so that the relationship between the calculated resonant tunneling energy $E_R$ (relative to the Fermi level) and measured $V_R$ becomes linear, and that $V_R$-$E_R$ goes through the origin; $V_R$=$sE_R$, where $s$ is a fitting parameter corresponding to the slope of $V_R$-$E_R$.[19,36,49,50] The $d$ values for the calculated data were slightly corrected so that the experimental data were reproduced well, where $r$ corresponds to the correction ratio of the etching rate estimated by this procedure to that estimated experimentally.

The color-coded $d^2I/dV^2$ intensities of samples A-G are shown in the top figures of Fig. 5(a)-(g), respectively. Here, these color-coded intensities are extrapolated from the measured data at the $d$ values corresponding to the white dots shown at the top of these figures. The calculated $V_R$ values of the HH and LH resonant levels as a function of $d$ are expressed by the connected violet and green dots, respectively. In all the samples, the calculated HH and LH resonant levels show good agreement with the measured data. In the bottom figures of Fig. 5(a)-(g), blue and red dots express the $s$ value used in the fitting and $RA$, respectively. The $RA$ value changes as a function of $d$, which can be attributed to the diffusion of the interstitial Mn atoms to the surface. Therefore, we changed $s$ gradually with increasing $d$. In most cases, we see a similar behavior between the $RA$-$d$ and $s$-$d$ curves, but their shapes are not perfectly the same. In our experiments, there is also a tunneling sequence where the holes injected in the VB quantum levels lose their energy and drop to IB by scattering in the (InGaMn)As (or InMnAs) QW. Therefore, the observed $RA$ values include the contribution of the conduction of the IB holes, whose details have not been clarified. On the other hand, $s$ is determined just by the energy positions of the VB resonant levels. These probably cause the different behavior between the $RA$-$d$ and $s$-$d$ curves. In the cases of (InGaMn)As devices (sample A-F), the resonant peaks observed in all of our samples are well reproduced by our model, which supports our conclusion that the resonant levels are formed by the quantization of the VB holes in (InGaMn)As layers. In the case of InMnAs (sample G), although the calculation can reproduce the experimental resonant peaks, the obtained $r$ value is much less than that in other samples and the higher resonant levels (LH1 and HH3) are not clearly observed in the experiments. The reason of these results is not clear, but it is probably due to the degradation of the crystallinity of InMnAs induced by the strong strain. In fact, the obtained $Q_\varepsilon$ (~0.53%) was much less than the expected value (~6.53%), which implies that the lattice of the InMnAs layer is almost relaxed. For more systematic and precise analysis, strain-free heterostructures are necessary for InMnAs, although it is very difficult to achieve them using the typical III-V materials.

In Fig. 6(a), the triangular red, circular blue, and rectangular green points show the $E_F$ values of $Ga_{1-x}Mn_xAs$,[37] $[(In_{0.53}Ga_{0.47})_{1-x}Mn_x]As$, and $In_{0.87}Mn_{0.13}As$ obtained in this study, respectively, where the values next to the data points are the $T_C$ values. The characters in the parentheses are the sample names. We see that $E_F$ decreases with increasing the In content, which can be attributed to the reduction of the binding energy of the Mn states with increasing the In content. In both GaMnAs and (InGaMn)As, $E_F$ decreases with increasing $T_C$ when $x$ is fixed, which can be explained by the increase of the IB hole concentration as shown in Fig. 6(b). Also, $E_F$ increases with increasing $x$, which can be qualitatively explained by the IB broadening as shown in Fig. 6(c). Similar behavior was theoretically expected in Ref. 51 and 52. These results are consistent with the IB conduction picture shown in Fig. 1(b). For more detailed discussions, we need to know further information of the precise IB characteristics which we cannot obtain by our method due to the orbital selectivity of the resonant tunneling spectroscopy.

**V. DISCUSSION**

Here, we show that one can rule out the possibility that these oscillations observed in Fig. 3(a)-(f) are induced by the quantized two-dimensional hole-gas states between the AlAs and InGaAs:Be.[53,54] We see that the observed results in Fig. 3(a)-(f) and (h) are typical features of the QW thickness dependence of resonant tunneling. For example, these oscillations clearly tend to be weakened with increasing $d$. This behavior can be understood by weakening of the quantization. Other essential feature is that the VB gradually branches out to sub-band levels with decreasing $d$. For instance, in Fig. 3(b) and (c), HH1 and LH1 (or HH2) levels are nearly merged at around 14-16 nm. These features are quite different from those of the quantized two-dimensional hole-gas states at AlAs/InGaAs:Be, if any. Also, if these oscillations were induced by quantized two-dimensional hole-gas states, the oscillation peak bias voltage would be proportional to $RA$ as mentioned in Ref. 53. However, as can be seen in Fig. 5(a)-(d), some oscillation peak voltages become smaller even when $RA$ increases with increasing $d$. Furthermore, as shown in our previous paper,[54] we did not see any clear oscillations induced by resonant tunneling in the $d^2I/dV^2$-$V$ curves of the tunnel device composed of GaMnAs/ AlAs (5nm)/ GaAs:Be ($1\times10^{18}$ cm$^{-3}$) when the surface GaMnAs thickness was 30 nm, where the quantization of the holes in GaMnAs is very weak. If the $d^2I/dV^2$ oscillations observed in our study were induced by the quantized two-dimensional hole-gas states, the oscillation would be observed even when the GaMnAs thickness is 30 nm. This result means that it is impossible to detect the quantized two-dimensional hole-gas states in our experimental setup when the Be doping level is $1\times10^{18}$ cm$^{-3}$.

The quantized two-dimensional hole-gas states formed at the AlAs/GaAs:Be interface have been observed when the Be concentration is extremely low (for example, $6\times10^{14}$ cm$^{-3}$ in Ref. 55). With increasing the Be concentration, however, the quantization of these states becomes weak. Also, as shown in Ref. 53, the energy separation between these states is only several meV when the Be concentration is $1\times10^{18}$ cm$^{-3}$. These are the reasons why we did not detect any quantized two-dimensional hole-gas states in our experiments. Also, we do not see any clear signals related to these states at AlAs/(GaAs:Be or a thin GaAs spacer) in the studies of the similar p-type GaAs-based resonant tunneling structures with the Be concentration of as high as $\sim10^{18}$ cm$^{-3}$.[19,50,56] Therefore, we conclude that these oscillations in Fig. 3(a)-(f) and (h) are attributed to the resonant tunneling in the (InGaMn)As (or InMnAs) QW layers.

## VI. SUMMARY

We performed the resonant tunneling spectroscopy on a series of [(In$_{0.53}$Ga$_{0.47}$)$_{1-x}$Mn$_x$]As films with $x$ from 6 to 15% and with $T_C$ from 70 to 135 K, and on the In$_{0.87}$Mn$_{0.13}$As film with $T_C$ of 47 K. Resonant tunneling was clearly observed in the (InGaMn)As and InMnAs films for the first time. In the cases of the (InGaMn)As films, by analyzing these resonant peak bias positions, we found that the almost unaffected VB of the host semiconductor InGaAs is still remaining in (InGaMn)As. In all the [(In$_{0.53}$Ga$_{0.47}$)$_{1-x}$Mn$_x$]As and In$_{0.87}$Mn$_{0.13}$As films investigated in this study, the p-d exchange splitting was not observed, which means that the p-d exchange splitting is negligibly small in VB. (probably several meV expected from our previous studies.[11,36]) The Fermi level exists in the band gap in all the [(In$_{0.53}$Ga$_{0.47}$)$_{1-x}$Mn$_x$]As and In$_{0.87}$Mn$_{0.13}$As samples. We found that $E_F$ decreases with increasing the In content, which can be attributed to the reduction of the binding energy of the Mn atoms at high In content. $E_F$ decreases with increasing $T_C$ when $x$ is fixed. Also, $E_F$ increases with increasing $x$. These results are consistent with the IB conduction picture. Our results strongly suggest that these are common features in (III,Mn)As (III=In and Ga). Our findings mean that the holes in VB are hardly related to the ferromagnetism, but the IB holes dominate the transport and magnetism of (III,Mn)As.

ACKNOWLEDGMENTS

This work was partly supported by Grants-in-Aids for Scientific Research, particularly Grant-in-Aid for Specially Promoted Research, the Special Coordination Program for Promoting Science and Technology, FIRST Program of the JSPS, PRESTO of the JST, and Asahi Glass Foundation.

TABLE I. Sample parameters, characteristics, and estimated $E_F$ of the $[(In_{0.53}Ga_{0.47})_{1-x}Mn_x]As$ (sample A-F) and InMnAs (sample G) resonant tunneling diodes investigated in this study. $a$ is the initial (InGaMn)As (or InMnAs) film thickness before etching. $T_S$ is the growth temperature. $T_A$ is the annealing temperature, and $t_A$ is the annealing time. We did not anneal sample E. The $T_C$ values were estimated from the Arrott plot using the data of the magnetic field dependence of magnetic circular dichroism or from the planar Hall effect measured for these samples before etching. $E_F$ is defined in the main text.

| Sample | $x$ (%) | $a$ (nm) | $T_S$ (°C) | $T_A$ (°C) | $t_A$ (hours) | $T_C$ (K) | $E_F$ (meV) |
|---|---|---|---|---|---|---|---|
| A (InGaMnAs) | 6 | 21 | 200 | 160 | 95 | 72 | 24 |
| B (InGaMnAs) | 9 | 20 | 190 | 160 | 25 | 70 | 40 |
| C (InGaMnAs) | 12.5 | 22 | 180 | 160 | 23 | 97 | 53 |
| D (InGaMnAs) | 15 | 23 | 180 | 160 | 16 | 115 | 62 |
| E (InGaMnAs) | 15 | 11.5 | 180 | --- | --- | 70 | 65 |
| F (InGaMnAs) | 15 | 11.5 | 180 | 160 | 5 | 135 | 55 |
| G (InMnAs) | 13 | 11.5 | 200 | 160 | 20 | 47 | 30 |

**Figure captions**

Figure 1 (a) General understanding of the density of states (DOSs) in (III,Mn)As (III=In and Ga) as a function of the electron energy. The top of the VB is largely deformed and merged with IB by Mn doping. The Fermi level, shown by red dash-dotted line, exists in the merged VB. (b) The schematic VB pictures of InMnAs and GaMnAs strongly suggested in our study, where the energy dependence of DOS and the wave vector $k_{x,y,z}$ dependence of the energy bands are schematically depicted in the left and right figures, respectively. The IBs of InMnAs and GaMnAs are expressed by the blue and orange areas sandwiched by the dotted curves (or lines) of each color, respectively. The Fermi levels of InMnAs and GaMnAs are expressed by the blue and orange dashed-dotted lines, respectively. Although we could not obtain clear evidence of the existence of IB, it is highly expected that it exists near the Fermi level. VB is not merged with IB and not largely affected by Mn doping. The Fermi level exists in IB in the band gap. The *p-d* exchange splitting is expected to be ~5 meV for the LH band.

Figure 2 (a) Schematic structures of our devices investigated in this study. The (InGaMn)As (or InMnAs) thickness $d$ (= $d_1$, $d_2$, and $d_3$) is varied by precise etching. (b) The VB profiles corresponding to the devices shown in (a). Here, orange (green) curves express the HH (LH) band when the quantum-size effect is neglected in the (InGaMn)As (or InMnAs) layers. The blue lines are the resonant levels. The red dash-dotted lines express the Fermi level. (c) The schematic graph of the ideal $d$ dependence of the resonant energy levels, where they converge at the VB top energy $E_V$.

Figure 3 (a)-(g) Black solid curves correspond to the $d^2I/dV^2$–$V$ characteristics of the $[(In_{0.53}Ga_{0.47})_{1-x}Mn_x]$As resonant tunneling devices of (a) sample A ($x$=6%, $T_C$: 72 K), (b) sample B ($x$=9%, $T_C$: 70 K), (c) sample C ($x$=12.5%, $T_C$: 97 K), (d) sample D ($x$=15%, $T_C$: 115 K), (e) sample E ($x$=15%, $T_C$: 70 K), (f) sample F ($x$=15%, $T_C$: 135 K), and (g) sample D ($x$=15%, $T_C$: 115K). The measurement temperature was 3.5 K. (h) Black solid curves correspond to the $d^2I/dV^2$–$V$ characteristics of the $In_{0.87}Mn_{0.13}$As resonant tunneling device (sample G) measured at 3.4 K. Note that the scale of -$V$ in (g) is different from those of (a)-(f) and (h). The dashed curves trace the positions of the resonant peaks. Here, HH$n$ and LH$n$ ($n$: 1,2,3…) represent resonant tunneling through the $n$th level of the HH and LH sub-bands in (InGaMn)As or InMnAs, respectively. Colors in these graphs express the $d^2I/dV^2$ intensity extrapolated from the measured data.

Figure 4 VB diagram of the (a) (InGaMn)As and (b) InMnAs resonant tunneling devices assumed in our calculations. Here, the thin green (black) and thick orange

(gray) lines express the VB of the LH and HH bands at the Γ point, respectively. The red dash-dotted line corresponds to the Fermi level.

Figure 5 (a)-(g) The upper graphs show the comparison between the calculated resonant levels and the experimentally obtained $d^2I/dV^2$ data of sample A-G, as functions of -$V$ and $d$. The $d^2I/dV^2$ intensity is expressed by color. Here, these color intensities are extrapolated from the measured data with $d$ corresponding to the white dots shown at the top of these figures. The connected violet and green dots are the calculated resonant peak bias voltages $V_R$ of the HH and LH bands, respectively. $E_F$, $r$, and $Q_\varepsilon$ are defined in the main text. In the lower graphs, blue and red dots express the $s$ value used in the fitting and the resistance area $RA$, respectively.

Figure 6 (a) Triangular red ,circular blue, and rectangular green dots are the $E_F$ values of GaMnAs, (InGaMn)As, and InMnAs respectively. The data of GaMnAs are those obtained in Ref. 37. The values next to the data points in this figure are the $T_C$ values. The characters in the parentheses are the sample names. (b) Schematic pictures showing how $E_F$ decreases with increasing $T_C$ or hole concentration when $x$ is fixed. (c) Schematic pictures showing how $E_F$ increases with increasing $x$. Here, CB means the conduction band.

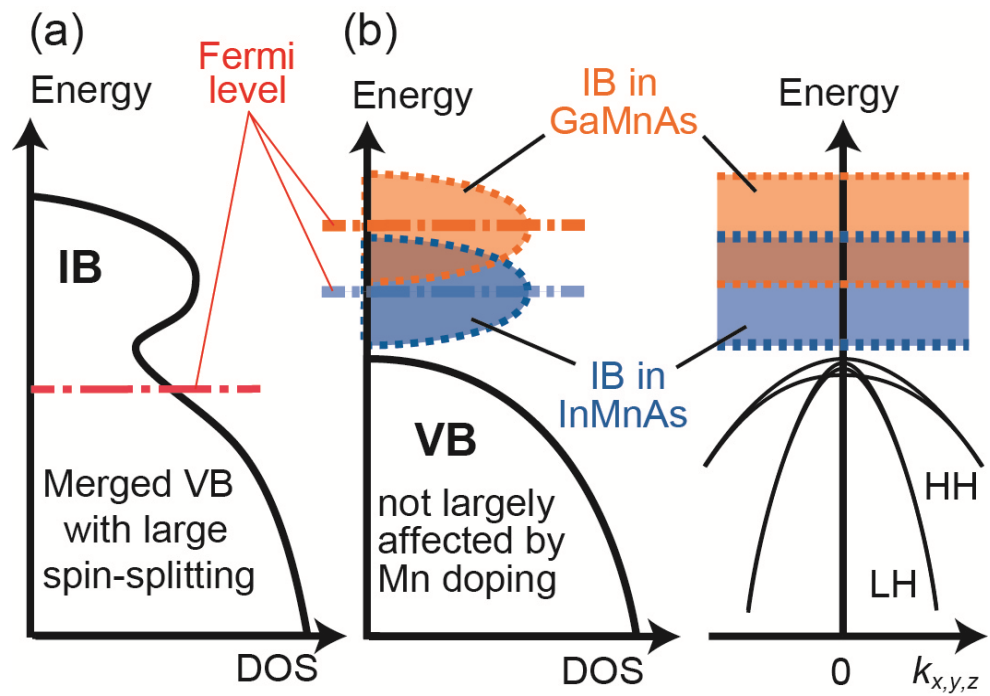

Figure 1

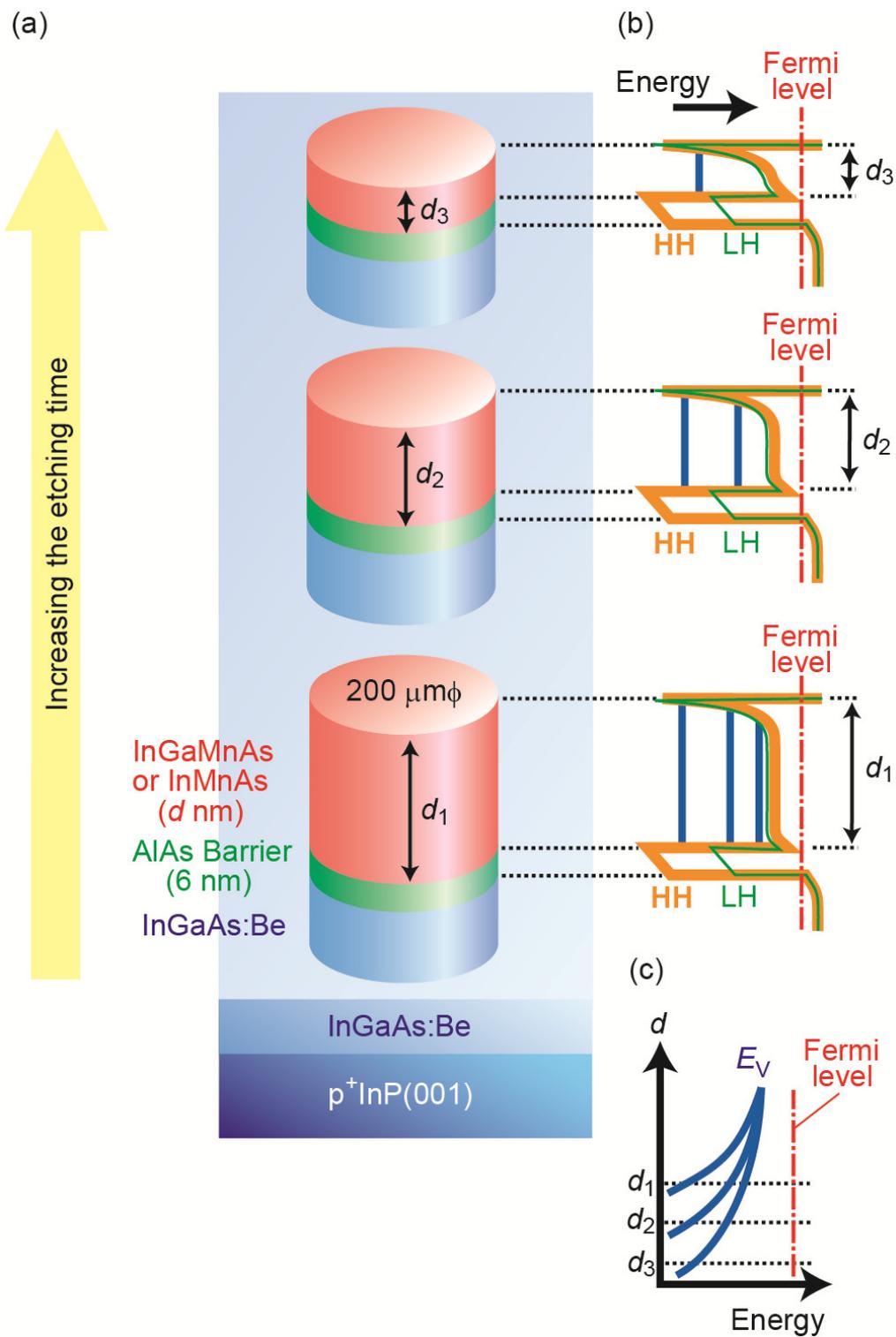

Figure 2

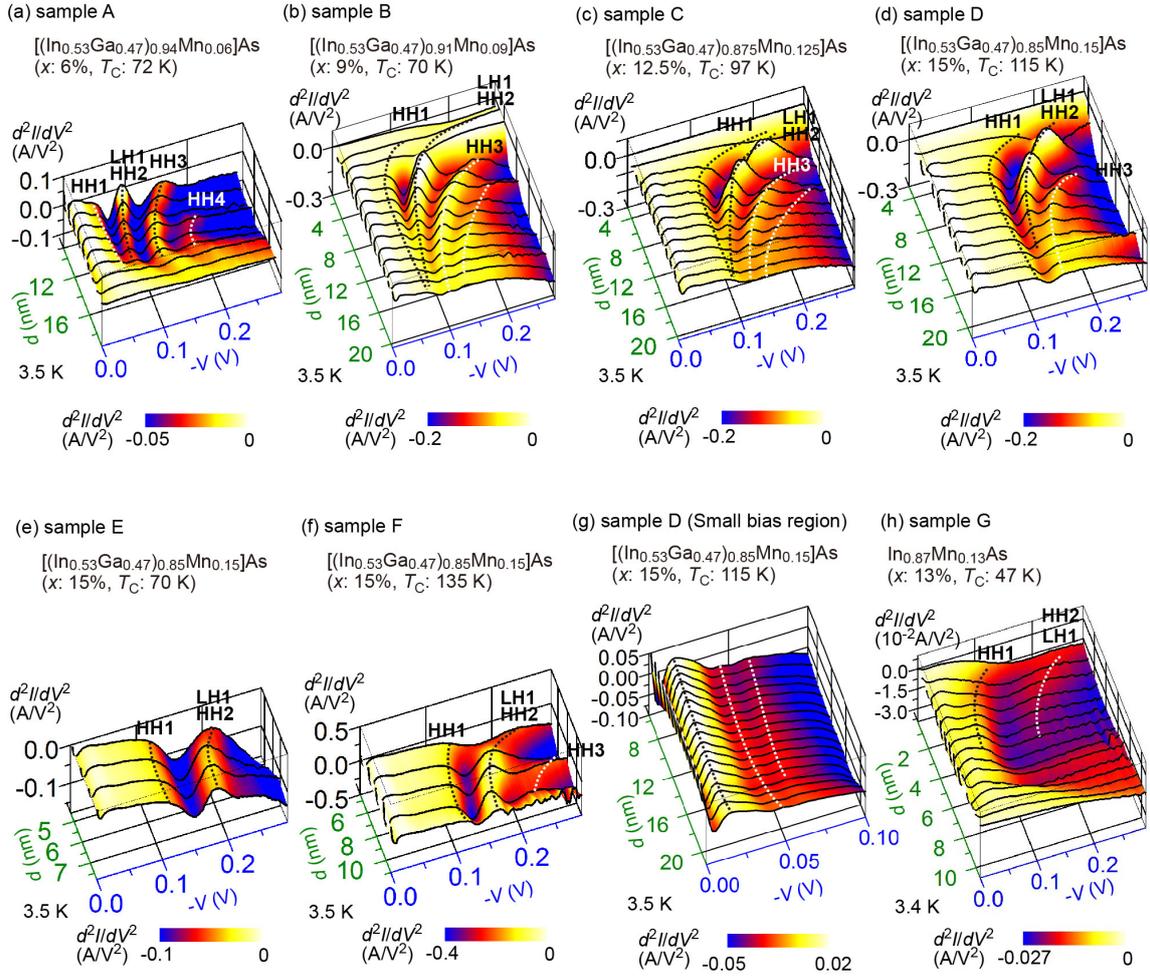

Figure 3

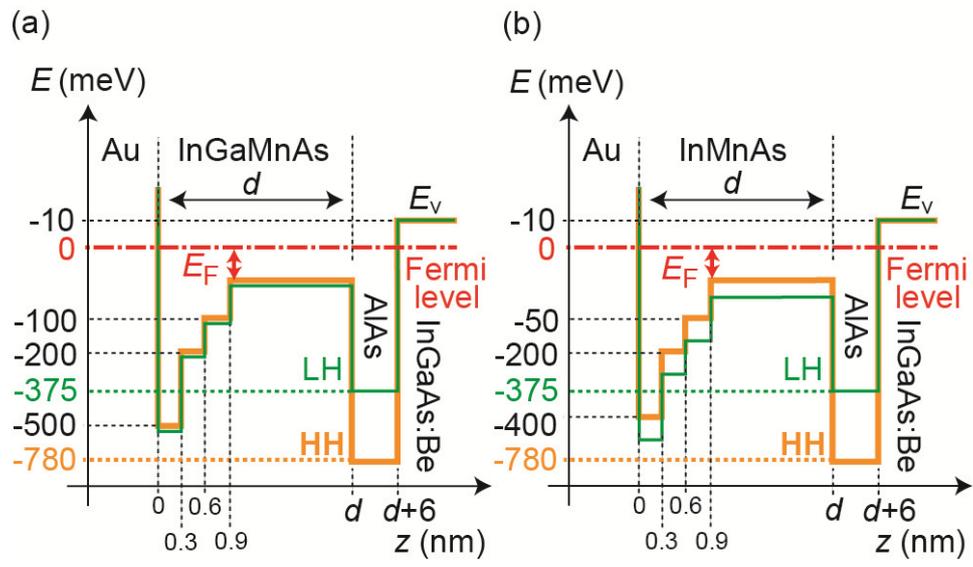

Figure 4

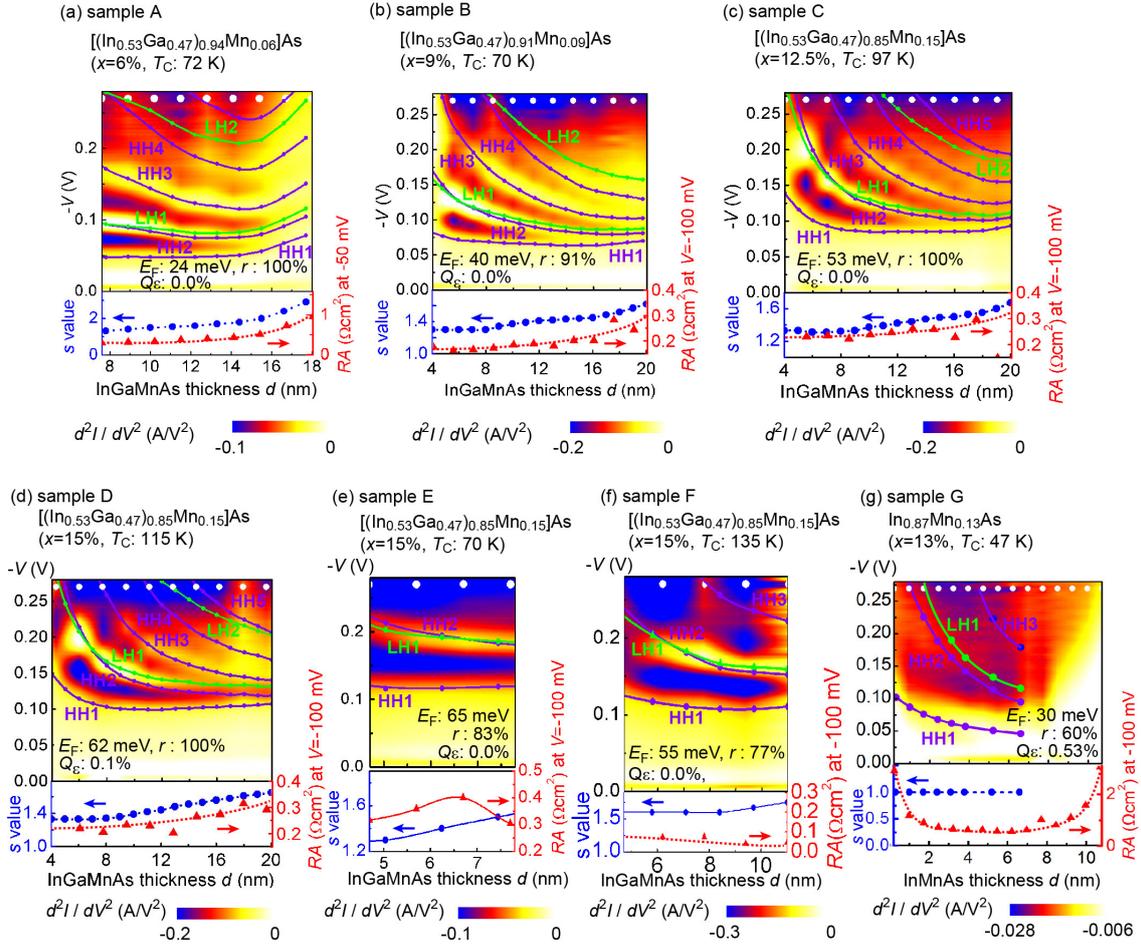

Figure 5

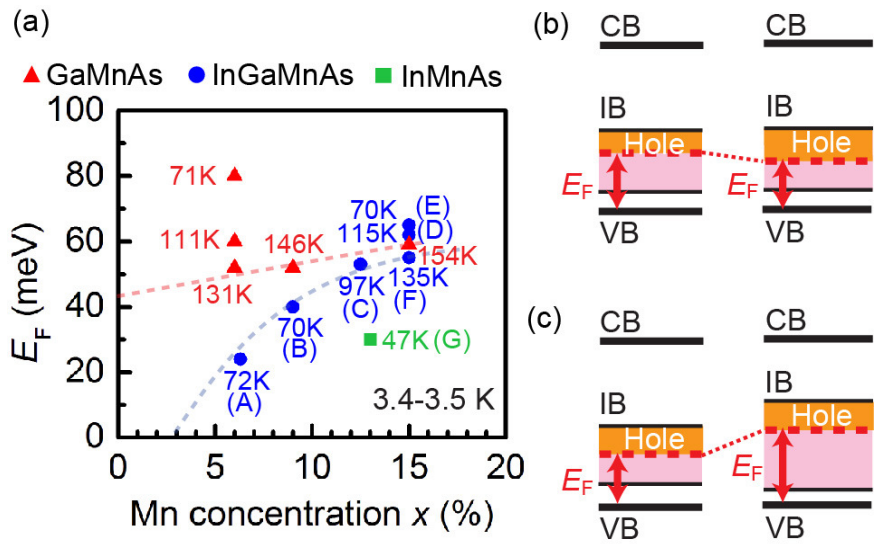

Figure 6